\documentclass[aps,prb,twocolumn,superscriptaddress,showpacs]{revtex4}

\usepackage{longtable}
\usepackage{graphicx}
\usepackage{dcolumn}
\usepackage{bm}
\usepackage{amsmath}
\usepackage[psamsfonts]{amssymb}
\usepackage{subfigure}

\newcommand{\BiSe}{Bi$_2$Se$_3$}

\newcommand{\MR}{magnetoresistance}
\newcommand{\ARPES}{angle resolved photoemission spectroscopy}
\newcommand{\STM}{scanning tunneling microscopy}

\begin{document}

\title{Probing the surface states in {\BiSe} using the Shubnikov-de Haas effect}



\author{M. Petrushevsky}
\affiliation{Raymond and Beverly Sackler School of Physics and Astronomy, Tel-Aviv
University, Tel Aviv, 69978, Israel}

\author{E. Lahoud}
\affiliation{Department of Physics, Technion-Israel Institute of Technology, Haifa 32000, Israel}

\author{A. Ron}
\affiliation{Raymond and Beverly Sackler School of Physics and Astronomy, Tel-Aviv
University, Tel Aviv, 69978, Israel}

\author{E. Maniv}
\affiliation{Raymond and Beverly Sackler School of Physics and Astronomy, Tel-Aviv
University, Tel Aviv, 69978, Israel}

\author{I. Diamant}
\affiliation{Raymond and Beverly Sackler School of Physics and Astronomy, Tel-Aviv
University, Tel Aviv, 69978, Israel}

\author{I. Neder}
\affiliation{Raymond and Beverly Sackler School of Physics and Astronomy, Tel-Aviv
University, Tel Aviv, 69978, Israel}

\author{S. Wiedmann}
\affiliation{High Field Magnet Laboratory, Institute for Molecules and Materials, Radboud University Nijmegen, Toernooiveld 7,
NL-6525 ED Nijmegen, The Netherlands}

\author{V. K. Guduru}
\affiliation{High Field Magnet Laboratory, Institute for Molecules and Materials, Radboud University Nijmegen, Toernooiveld 7,
NL-6525 ED Nijmegen, The Netherlands}

\author{F. Chiappini}
\affiliation{High Field Magnet Laboratory, Institute for Molecules and Materials, Radboud University Nijmegen, Toernooiveld 7,
NL-6525 ED Nijmegen, The Netherlands}

\author{U. Zeitler}
\affiliation{High Field Magnet Laboratory, Institute for Molecules and Materials, Radboud University Nijmegen, Toernooiveld 7,
NL-6525 ED Nijmegen, The Netherlands}

\author{J. C. Maan}
\affiliation{High Field Magnet Laboratory, Institute for Molecules and Materials, Radboud University Nijmegen, Toernooiveld 7,
NL-6525 ED Nijmegen, The Netherlands}

\author{K. Chashka}
\affiliation{Department of Physics, Technion-Israel Institute of Technology, Haifa 32000, Israel}

\author{A. Kanigel}
\affiliation{Department of Physics, Technion-Israel Institute of Technology, Haifa 32000, Israel}

\author{Y. Dagan}
\email[]{yodagan@post.tau.ac.il}\affiliation{Raymond and Beverly Sackler School of Physics and Astronomy, Tel-Aviv University, Tel Aviv, 69978,
Israel}


\date{\today}

\begin{abstract}
Shubnikov-de Haas (SdH) oscillations are observed in {\BiSe} flakes with high carrier concentration and low bulk mobility. These oscillations probe the protected surface states and enable us to extract their carrier concentration, effective mass and Dingle temperature. The Fermi momentum obtained is in agreement with {\ARPES} (ARPES) measurements performed on crystals from the same batch. We study the behavior of the Berry phase as a function of magnetic field. The standard theoretical considerations fail to explain the observed behavior.

\end{abstract}



\maketitle
\section{Introduction}
The recently discovered topological insulators are a matter of intense theoretical and experimental study in contemporary condensed matter physics (for a review of the field, see ref. \cite{HasanReview}). The exotic surface states that occur in these materials were confirmed and thoroughly studied using surface sensitive techniques, such as ARPES \cite{BiSeARPES} and {\STM} (STM) \cite{BiSbSTM}. These measurements demonstrated the absence of backscattering at the surface, which is believed to be protected by time reversal symmetry.

\par
Of the 3D topological insulators, {\BiSe} has the simplest band structure - a single surface Dirac cone, however transport experiments on this material present a challenge. The surface conducting state in {\BiSe} is difficult to separate from the spurious bulk conductivity, believed to originate from Se vacancies \cite{Bi2Se31974,BiSeControlledDoping,FisherBulkBiSe}. Even in the lowest carrier concentration samples metallic bulk behavior is observed. A few successful attempts to surmount this problem were reported, including: thinning down the 3D material to decrease the bulk contribution \cite{NanoribbonsAB,TangNanoRibbons,BiSeMBE,BiSeThinFilms}, using gate voltage \cite{LuGateVoltage,OngVoltageTunedBiSe}, pulsed high magnetic fields \cite{Analytis55T}, and changing the material composition to reduce the bulk conductivity \cite{BiSeTunable,BiSeControlledDoping,AndoBiTeSe}. However, providing conclusive transport evidence of the $\pi$ Berry phase expected from the topology of the surface remains a challenge.

\par
Recent attempts to probe the surface of {\BiSe} crystals using the SdH effect resulted in an overwhelming bulk signal \cite{FisherBulkBiSe,Paglione,AndoBulkBiSe}. These groups focused on making the samples cleaner in order to increase the bulk resistance. However, the decrease in the bulk carrier concentration led to an increased mobility. Consequently, the 3D oscillations dominated the signal.

\begin{figure}
\begin{center}
\includegraphics[width=1\hsize]{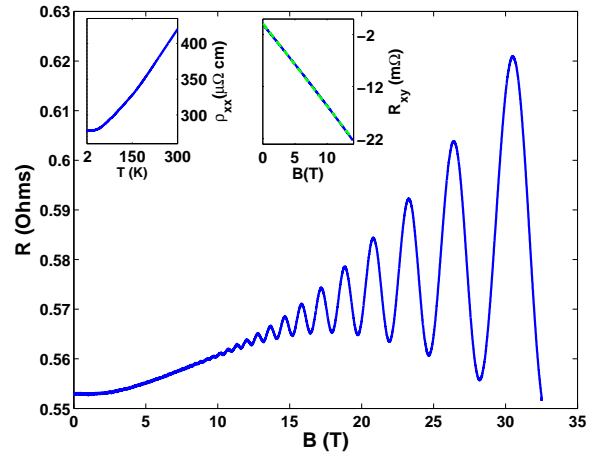}
\caption {(Color online) Resistance as a function of field for a high carrier concentration (S1) at $T=4.2K$ in a perpendicular magnetic field up to 32.5T. Left inset: typical longitudinal resistance versus temperature at $H=0$. Right inset: Hall resistance as a function of magnetic field at $T=2K$, $H \parallel C_3$ axis. The dashed line is a linear fit yielding a charge carrier concentration of $5.6\times10^{19}cm^{-3}$. From the sign of the Hall slope we find that the charge carriers are electrons.
\label{fig:BasicInfo}}
\end{center}
\end{figure}

\par
In our work we adopt the opposite approach; the SdH effect has an exponential sensitivity on the mobility, whereas the conductance varies only linearly with it. This allows a mobility window in which the SdH effect probes only the protected surface states, albeit the dominance of the bulk over the classical conductance. This can be done by increasing the carrier concentration (thus lowering the bulk mobility), in contrast to the conventional approach.
\par
Here we report on transport and ARPES measurements in {\BiSe} flakes. We observe SdH oscillations in the {\MR}, and provide evidence that they originate solely from the surface Dirac cone. From these measurements we find the effective mass and carrier concentration and provide a lower bound for the Dingle scattering time. The ARPES data provides evidence of a single Dirac cone and the Fermi wave vector of the surface states obtained is in agreement with the magneto transport result. We study the evolution of the SdH phase with magnetic field. The broad field range studied and the pronounced oscillations observed result in unprecedented accuracy at which the frequency can be determined. However, we find that the phase is a strong function of the frequency. The data is consistent with a $\pi$ Berry phase changing with magnetic field. However, the conventional Zeeman effect alone is insufficient to describe our data. Other possibilities, such as zero phase, are discussed.
\section{Methods}
\par
Single crystals of {\BiSe} were grown in the standard method \cite{Bi2Se31974,HorSuperconductivity}. It is known that a Se rich mixture usually results in a reduced carrier concentration in the bulk, as is the case in our reference sample with a carrier concentration of $\sim10^{17} cm^{-3}$. In order to get a rather high carrier concentration and low bulk mobility, we used a stoichiometric mixture of Bi (99.999$\%$) and Se (99.99$\%$), which was cooled down slowly at 3$^{\circ}C/hour$ and annealed for 24h at 620$^{\circ}C$. This resulted in a carrier concentration of $\sim10^{19} cm^{-3}$. X-ray diffraction spectra (not shown) are consistent with previous reports of single crystals of {\BiSe} \cite{AndoBulkBiSe}.

\par
Flakes $\sim10-100\mu m$ thick were freshly cleaved perpendicular to the $C_3$ axis from a single crystal in a nitrogen environment (samples S1-S5). Gold contact wires were attached to the sample using silver paint. Measurements up to 14T were performed in a Quantum Design PPMS platform using Keithley 6221 current-source coupled to a Keithley nanovoltmeter 2182A. High field measurements (up to 32.5T) were performed in the HFML with an SR830 DSP Lock-In Amplifier. For SdH measurements flakes were cut into narrow bars and measured in a four contact configuration. Other pieces from the same batch were measured in a 5-contact and Van der Pauw configurations for a more accurate determination of the resistivity and Hall coefficient. Two Hall probes were used for accurate determination of the perpendicular and parallel field components.
ARPES measurements were performed using the HeI$^\alpha$-line (h$\nu$=21.2eV) from a Scienta UV-lamp and with an R4000 Scienta analyzer. The crystals for the ARPES measurement were cleaved in-situ at vacuum better than 5x10$^{-11}$torr at 20K.
\par
We took extra care to properly align our voltage leads. Misaligned contacts resulted in a spurious Hall contribution in the longitudinal voltage channel. This contribution manifested itself in an antisymmetric background and a phase shift between positive and negative magnetic fields. Although all samples gave similar frequencies, samples with such Hall contribution were not used for the phase analysis.

\section{Results and discussion}
\par
In the insets of Figure \ref{fig:BasicInfo} we show typical transport properties of a sample with high carrier concentration.
The resistivity as a function of temperature for a typical flake is shown in the left inset. A metallic temperature dependance with a saturation value of $\sim280\mu\Omega cm$ below 20K is observed. The carrier concentration of this flake inferred from the Hall resistance at 2K (see right inset) is $5.6\times10^{19}cm^{-3}$, this gives a bulk mobility of $400 cm^2/V\cdot sec$. Other samples in the literature have lower carrier density and higher mobilities \cite{FisherBulkBiSe, Paglione}. These studies also show samples with higher carrier concentration $\sim 10^{19} cm^{-3}$, however, their bulk mobility is at least a factor of two higher. Apparently, all these reported mobilities are high enough to obscure surface oscillations. Moreover, our reference (non stoichiometric) samples with $n\sim10^{17} cm^{-3}$ have similar properties as other samples in the literature \cite{Paglione} exhibiting 3D oscillations (see Appendix D).

\par
The resistance versus magnetic field at 4.2K is shown in Fig. \ref{fig:BasicInfo}.
Strong oscillations are observed. Below we analyze the oscillations and show evidence that they arise from the topological surface states.

\begin{figure}
\begin{center}
\includegraphics[width=1\hsize]{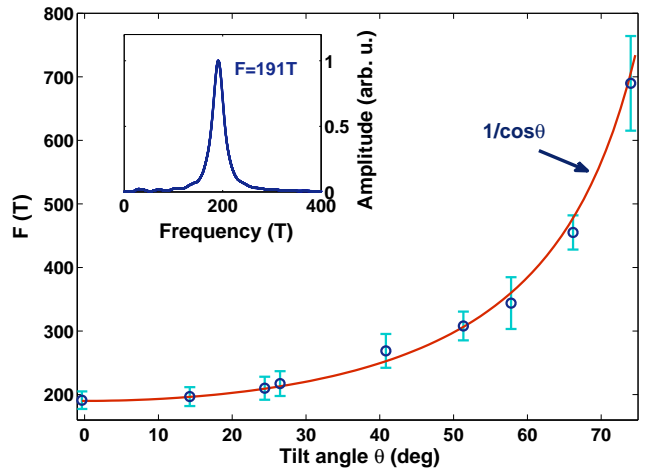}
\caption {(Color online) (a) Fourier transform (FFT) of the oscillations in S2 (similar to S1) after substracting a smooth polynomial background. (b) Frequency of the oscillations measured in S2 as a function of $\theta$ (all other samples from this batch exhibited similar behavior). The bars represent the Full Width at Half Maximum of the FFTs, which are an upper bound for the error. The solid line is the expected 2D $\frac{1}{cos \theta}$ behavior. \label{fig:SdHOsc}}
\end{center}
\end{figure}

\par
According to the Onsager relation \cite{SdH},
\begin{equation}
F=\frac{\hbar}{2\pi e}A(\epsilon_F),
\label{eq:Onsager}
\end{equation}

the frequency of the SdH oscillations $F$ is proportional to $A(\epsilon_F)$, the cross section of the Fermi surface in the plane perpendicular to the applied magnetic field. $\hbar$ is Planck's constant and $e$ is the electron charge. For the case of a two-dimensional (2D) Fermi surface, $F\propto\frac{1}{cos\theta} $, where $\theta$ is the angle between the normal to the 2D plane and the direction of the magnetic field. In Figure \ref{fig:SdHOsc} we show that our data follows the expected behavior of a 2D Fermi surface. Here $\theta$ is the angle between the $C_{3}$ axis and the applied magnetic field. For $\theta=0$, we find frequencies in the range of $190-198T$ for S1-S5.

\par
The behavior shown in Figure \ref{fig:SdHOsc} is in strong contrast with measurements of our low carrier concentration samples. For these flakes, $F=20,26T$ for $H \parallel C_{3}$ and $H \perp C_{3}$, respectively. This corresponds to a 3D carrier density of $6.6\times10^{17} cm^{-3}$ \cite{Kulbachinskii1999}, consistent with Hall measurements (Data are presented in Appendix D).

 \par
Substituting the average frequency $F=195T$ in Eq. \ref{eq:Onsager} and assuming a circular Fermi surface, we find $k_F = 0.077\pm0.003{\AA}^{-1}$, which corresponds to a 2D carrier density of $n_2$$_D = \frac{k_F^{2}}{4\pi}=4.7\times10^{12} cm^{-2}$, assuming no spin degeneracy, as should be the case for a topologically protected surface.

 \par
\begin{figure}
\begin{center}
\includegraphics[width=1.\hsize]{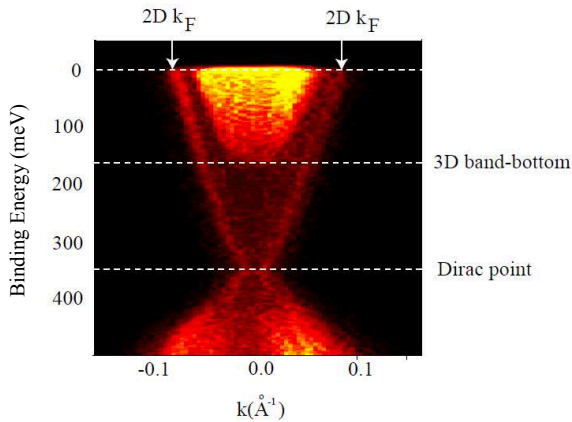}
\caption {ARPES band dispersion of a high carrier concentration sample. Both the bulk and the surface states are observed in the data. The dashed lines mark the bottom of the 3D band (around
160meV below the Fermi-level) and the Dirac-point (around 345meV below the
Fermi-level). The arrows mark k$_F$ of the 2D surface state
($k_F = 0.084 \pm 0.005 {\AA}^{-1}$).\label{fig:ARPES}}
\end{center}
\end{figure}

Fig. \ref{fig:ARPES} presents ARPES data from a cut taken along the $\Gamma$M direction and going through the $\Gamma$ point on a typical flake. The contribution of the bulk-electrons as well as the Dirac-like surface state are clearly seen. We find k$_F=0.084 \pm 0.005{\AA}^{-1}$ in agreement with the value extracted from the SdH analysis.

\par
The resistance of a 2D system exhibiting SdH oscillations is given by\cite{SdH}
\begin{equation}
{R_s} ^{2D}=R_0\{1+R_TR_Dcos[2\pi(\frac{F}{B}+\frac{1}{2}+\beta)]\}
\label{eq:Sdh1}
\end{equation}
where $R_0$ is the zero field resistance and $2\pi\beta$ is the Berry phase which is expected to be $\pi$ for an electron rotating around the Dirac point in a topological insulator at the low magnetic field limit \cite{HasanReview}. 
$R_T$ contains the temperature dependance,
\begin{equation}
R_T=\frac{\alpha T}{B}/sinh(\frac{\alpha T}{B})
\label{eq:Sdh2}
\end{equation}
with $\alpha = \frac{2\pi^{2}m^{*}k_B}{\hbar e}$. The dingle factor is
\begin{equation}
R_D = exp(\frac{-\pi}{\omega_c \tau_D})
\label{eq:Sdh3}
\end{equation}
where $\tau_D$ is the dingle scattering time, and corresponds to the dephasing of the Landau states. $\omega_c=\frac{e B}{m^{*}}$ is the cyclotron frequency. $R_D$ determines the amplitude decay with the decrease of magnetic field.

\par
\begin{figure}
\begin{center}
{\includegraphics[width=1\hsize]{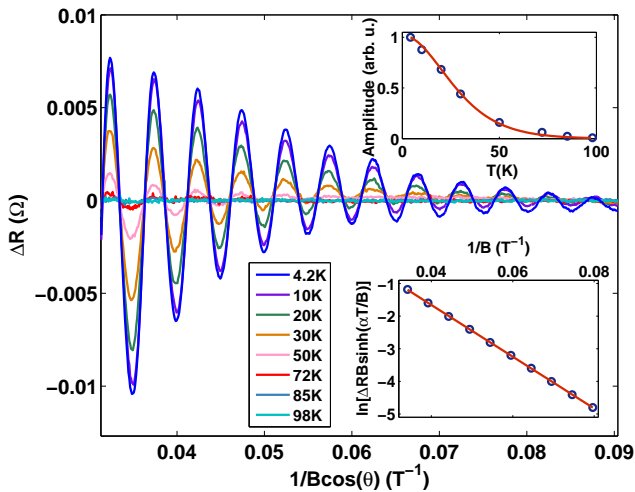}}
\caption {(Color online) Temperature dependence of the SdH oscillations at $\theta=0$ after subtracting a smooth polynomial background (S3). Top inset shows the temperature dependence of the amplitude, yielding m*$\simeq$$0.16m$$_e$. Bottom inset: a Dingle plot resulting in $\tau _{D} = (3.6\pm0.4)\times10^{-14}sec$.\label{fig:Temp}}
\end{center}
\end{figure}

The standard analysis of SdH oscillations yields a cyclotron mass of $m^{*}\simeq0.16m_e$ \footnote{The cyclotron mass should not be zero even for the case of a linear dispersion relation. It is determined from the relation $m^{*}=E_F / v_F^{2}$ \cite{Graphene}.}  (see Fig. \ref{fig:Temp}), this together with the field dependance of the oscillation amplitude at 4.2K gives $\tau_D = (3.6\pm0.4)\times10^{-14}sec$, a Dingle temperature of $T_D = \frac{\hbar}{2\pi\tau_D k_B} = 33.78\pm 0.33K$ and a corresponding quantum mobility of $\mu_{q}^{2D}=400 cm^2/V\cdot sec$.

\par
Further confirmation of the 2D nature of the oscillations reported here is inferred from a comparison between the quantum (Dingle) and transport mobilities. Forcing a 3D fit to the amplitude of the oscillations (which contains an additional $\sqrt{B}$ coefficient), one finds a 3D Dingle time of $3.9\times10^{-14}sec$ which corresponds to a bulk quantum mobility of $\mu_{q}^{3D}=570 cm^2/V\cdot sec$, larger than the transport mobility $\mu_{tr}^{3D}=400 cm^2/V\cdot sec$. This is in strong contrast with recent reports for {\BiSe}, where the Dingle mobility is four times smaller than the transport one \cite{AndoBulkBiSe}. Furthermore, it is physically impossible to have $\mu_{q}^{3D} > \mu_{tr}^{3D}$ in a bulk material, since the Dingle mobility takes into account all scattering events including small angle scattering that usually do not affect the resistivity, whereas for the transport mobility, backscattering events play the major role.

\par
One of the hallmarks of a topological insulator is the $\pi$ Berry phase associated with the Dirac dispersion relation. Since pronounced oscillations are observed in a broad field range (see Figure \ref{fig:BasicInfo}), we are able to follow the evolution of the Berry phase as a function of applied magnetic field for S1.
\par
The standard phase analysis is done by plotting $F/B$ versus $n$, the oscillation index (usually referred to as the Landau level fan diagram). Applying this to our data, using the FFT peak $F=198T$ yields a zero phase ($\beta=0$). We note that due to the many oscillations measured, the Fourier transform analysis favors a frequency that yields a zero phase. Furthermore, we find that $\beta$ is very sensitive to F, and even a deviation of $1.5\%$, which is well within the error margin, has a strong effect on $\beta$. For example, if we take $F=195T$, we get a $\pi$ Berry phase ($\beta=0.5$) at low fields, deviating towards zero as the field increases. In the following, we study the behavior of $\beta$ for frequencies in the vicinity of the FFT peak, $F=198\pm 3T$.

\begin{figure}
\begin{center}
{\includegraphics[width=1\hsize]{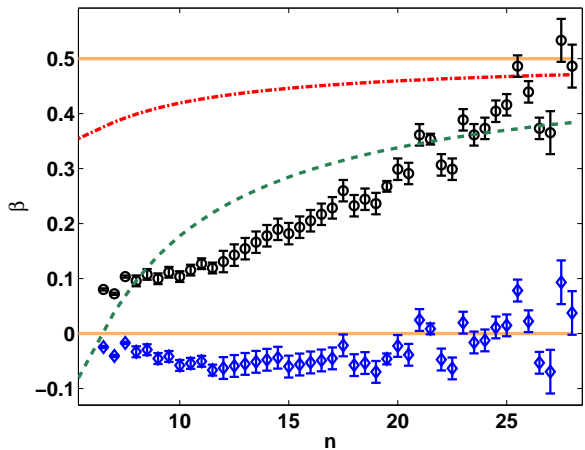}}
\caption {(Color online) Analysis of the Berry phase $2 \pi \beta$ according to Eq. \ref{eq:beta} using the data from Figure \ref{fig:BasicInfo} after subtracting a smooth polynomial background. $n=6.5$ corresponds to $B=30.4$T and $n=28$ to $B=7$T. The blue diamonds are obtained using the FFT peak of $F=198$T, consistent with a zero Berry phase. The black circles correspond to $F=195$T, for which a $\pi$ Berry phase is observed at low fields. This phase changes towards zero as the field increases. This demonstrates the strong sensitivity of the phase to the frequency chosen. The error bars stem from the freedom in choosing the oscillation extrema. The green dashed (red dashed-dot) line is the expected behavior of $\beta$ taking into account a Zeeman term with $g=50$ ($100$) respectively.\label{fig:LLfanDiagram}}
\end{center}
\end{figure}

\par
In order to focus on the field and frequency dependencies of the phase, we use the relation
\begin{equation}
\beta(F,n)=n-F/B
\label{eq:beta}
\end{equation}
and plot $\beta$ versus $n$ in Figure \ref{fig:LLfanDiagram}, where $n$ is the index of the oscillation minima for the two frequencies mentioned (see Appendixes A and B for more details and analysis on more samples). The two behaviors mentioned above are now clearly seen. This analysis casts strong doubts on the ability to independently determine the Berry phase using SdH oscillations.

\par
Assuming that at low field $\beta$ should be 0.5, the simplest explanation for the phase change would be a Zeeman term in the Hamiltonian \cite{Analytis55T}. In Figure \ref{fig:LLfanDiagram} we plot the expected behavior of $\beta$ using gyromagnetic ratios $g=50,100$ and $v_{F}=6.24\times10^{5}m/sec$ as inferred from the ARPES measurement (see Figure \ref{fig:ARPES}). While both theoretical curve ($g=100$) and the measured data are in qualitative agreement, it is obvious that this simple model is insufficient to describe our measurements. It is possible that the non-ideality of the Dirac dispersion should also be taken into account \cite{AndoBerryPhase}.

\section{Conclusion}
\par
In summary, we present evidence that the surface states in {\BiSe} can be probed in highly conducting flakes using the Shubnikov-de Haas (SdH) effect. This is in contrast with the conventional approach focused on improving the crystal purity. From the SdH analysis we find for the surface states $m^{*}=0.16m_e$, $\tau_{D}=(3.6\pm0.4)\times10^{-14}sec$. The Fermi momentum is in agreement with the ARPES data obtained on a flake from the same batch. We carefully study the behavior of the Berry phase with magnetic field, and show that two scenarios are possible: $\beta = 0.5$ ($\pi$ Berry phase) at low magnetic fields which changes fairly quickly with magnetic field, or a trivial zero phase. .
Two major issues are yet to be understood: first, the survival of protected surface states despite the large bulk conductivity and second, the peculiar behavior of the Berry phase with magnetic field.

\begin{acknowledgments}
 The work at Tel Aviv University is supported by the Israel Science Foundation (ISF) under grant No. 1421/08 and the Ministry of Science and Technology. Part of this work has been supported by EuroMagNET under the EU contract $n^{\circ} 228043$.

\end{acknowledgments}

\appendix
\section{Zeeman effect on $\beta\left(n\right)$}
We use a simplified model of non-interacting electrons on the surface of topological insulators
in a perpendicular external magnetic field. These electrons occupy orbital Landau
levels which are coupled to their spin. The spin is also Zeeman coupled to the magnetic field. The energy of an electron in the $n$'th orbital
Landau level is therefore composed from both the orbital energy and
the Zeeman energy in the following way (see for example Ref. \cite{AndoBerryPhase}):

\begin{equation}
E_{n}=\sqrt{\frac{1}{4}\left(g\mu_{B}B\right)^{2}+2n\hbar^{2}\left(\frac{v_{F}}{l_{b}}\right)^{2}},\label{eq:E_n}\end{equation}where $g$ is the effective gyromagnetic ratio for electrons on the surface,
$\mu_{B}$ is the Bohr magneton, $B$ is the magnetic field, $v_{F}$
is the Fermi velocity associated with the Dirac cone, and
$l_{b}=\sqrt{\frac{\hbar}{eB}}$ is the magnetic length. The $n$'th
maximum of the SdH oscillation in $R_{xx}$ occurs at magnetic field
$B_{n,max}$ where this energy crosses the Fermi energy, $E_{n}=E_{F}$.
This requirement and Eq.~(\ref{eq:E_n}) lead to

\begin{equation}
GB_{n,max}+n=\frac{F_{0}}{B_{n,max}}\label{eq: G,F max}\end{equation}where $G=\frac{1}{8e\hbar}\left(\frac{g^{*}\mu_{B}}{v_{F}}\right)^{2}$,
and $F_{0}=E_{F}^{2}/2\hbar ev_{F}^{2}$. One can write similar expression for the minimum of the oscillations
which occurs half way between the $n$ and $n+1$ maxima. Interpolation of Eq.~(\ref{eq: G,F max})
gives approximately (for large enough values of $n$),\[
GB_{n,min}+n+\frac{1}{2} \simeq\frac{F_{0}}{B_{n,min}}.\]

The phase of the oscillations is defined as in Eq.~(5) in the paper,
\begin{equation}
\beta\left(F,n\right)=n+\frac{1}{2}-\frac{F}{B_{n,max}}\simeq n-\frac{F}{B_{n,min}}\label{eq: G,F min}\end{equation}
In usual metals, this definition gives $\beta=0$, independent of $n$, for the right choice of $F$. However, in our case, given the theoretical relation between $n$ and $B_{n,max(min)}$ in Eq.~(\ref{eq: G,F max}) and (\ref{eq: G,F min}),
$\beta$ depends on $n$ for any choice of $F$. We can choose $F=F_{0}$,
which is the frequency of the oscillations in the limit of low magnetic
fields, and we get for $n\gg1,$

\begin{eqnarray*}
\beta(F_{0},n) & = & 0.5+GB_{n,max}\simeq GB_{n,min}\end{eqnarray*}

\begin{figure}
\begin{center}
\includegraphics[width=0.5\textwidth,height=0.28\textheight]{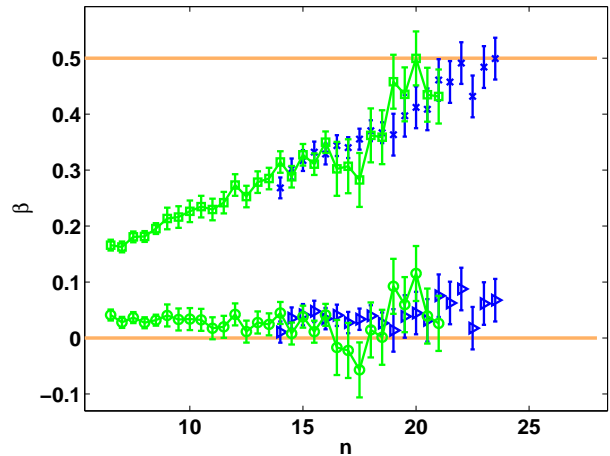}
\caption {(Color online) The behavior of $\beta(n)$ for S4 (green circles - $196T$ and squares - $192.8T$) and S5 (blue Xes - $186.5T$ and triangles - $190T$). \label{fig:Berry2}}
\end{center}
\end{figure}

\section{Frequency dependence of $\beta\left(n\right)$ for additional samples}

\par
In Fig. \ref{fig:Berry2} we show the frequency dependence of $\beta(n)$ for additional samples -  S4 and S5 (measured up to 33T and 14T, respectfully). One can see that this is similar to the behavior of S1 depicted in Fig. \ref{fig:LLfanDiagram}.

\section{Bulk {\MR}}

\par
\begin{figure}
\begin{center}
\includegraphics[width=0.5\textwidth,height=0.28\textheight]{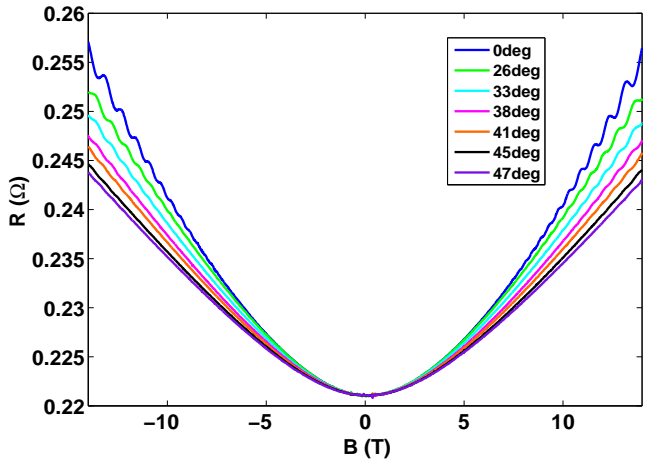}
\caption {(Color online) Resistance as a function of magnetic field for various tilt angles $\theta$ (S5).\label{fig:MR}}
\end{center}
\end{figure}

One may conjecture that the apparent 2D angular behavior is a result of some anomalous scattering occurring in the bulk when applying a parallel magnetic field component. To refute this claim we show in Figure \ref{fig:MR} the angular dependence of the {\MR}. One can see that the {\MR} decreases with $\theta$. This implies that the bulk scattering time increases with tilt angle. Furthermore, the {\MR} as a function of perpendicular magnetic field is not constant, but slightly increasing. Thus a quasi-2D bulk is not likely as well.

\section{Lower carrier concentration samples}

\begin{figure}
\begin{center}
\includegraphics[width=0.5\textwidth,height=0.28\textheight]{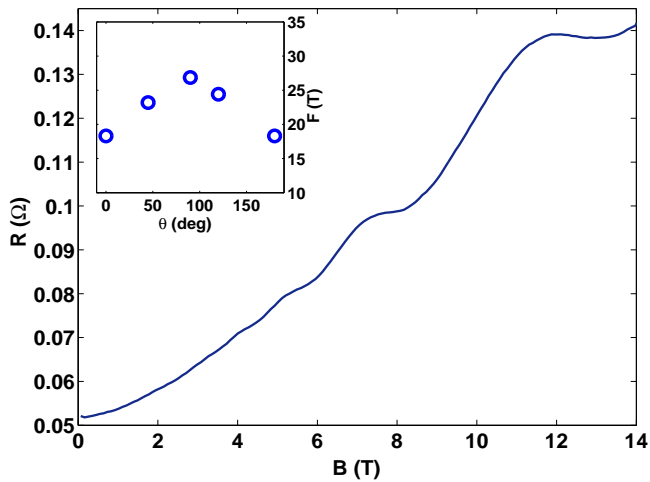}
\caption {(Color online) Typical Oscillations in our lower carrier concentration samples of $n\sim10^{17} cm^{-3}$. The inset is the Typical 3D angular dependence of the oscillations frequency.\label{fig:3DSample}}
\end{center}
\end{figure}

In Fig. \ref{fig:3DSample} we demonstrate the typical 3D behavior observed in our reference, lower carrier concentration samples with $n\sim10^{17} cm^{-3}$. These samples were prepared with a Se concentration higher than the stoichiometric one. This is done in order to reduce the number of Se vacancies and consequently the bulk conductance resulting in an increased mobility. The observed behavior for these samples is consistent with \cite{Paglione}.

\bibliographystyle{apsrev}
\bibliography{myBib}
\end{document}